\documentclass[reprint,nofootinbib,amsmath,amssymb,aps]{revtex4-1}
\usepackage{graphicx}

\newcommand{\be}{\begin{equation}}
\newcommand{\ee}{\end{equation}}
\newcommand{\bea}{\begin{eqnarray}}
\newcommand{\eea}{\end{eqnarray}}
\newcommand{\bes}{\begin{subequations}}
\newcommand{\ees}{\end{subequations}}
\newcommand{\bear}{\begin{equation}\begin{array}}
\newcommand{\eear}[1]{\end{array}\label{#1}\end{equation}}
\newcommand{\fr}[2]{\dfrac{{ #1}}{{ #2}}}

\newcommand{\fn}[1]{\footnote{{\sf #1}}}
\newcommand{\bu}{$\bullet$\ }

\begin{document}
\title{Simple and robust method  for search Dark Matter particles and measuring  their properties  at ILC in various models of DM}
\author{I. F. Ginzburg}
\affiliation{ Sobolev Institute of Mathematics and Novosibirsk State University\\
{\it Novosibirsk, Russia}}


\begin{abstract}
{I suggest simple  method for the search   of Dark Matter particles and some related particles which allows to measure reliably their mass and spin in a wide class of models for Dark Matter.
}

\end{abstract}

\maketitle
\section{Introduction}\label{secintro}

About 25\% of the Universe is made from  Dark Matter (DM). There are a number of models in which DM consists from particles similar to those in SM with additional discrete quantum number, which I denote here as D-parity.  For known particles $D=1$, for new DM particles (DMP) $D=-1$, and D-parity conservation ensures stability of lightest from these new particles.   We will consider such  models for DM, in which there is more than one particle with parity $D=-1$, in particular, neutral DMP, denoted below as D, and its more heavy lightest charged counterpart $D^\pm$. It is assumed that all these D-particles have identical spin $s_D$   (1/2 or 0).

\bu The well known examples of such models for DM are given by MSSM and NMSSM. Here D-parity is another name for R-parity, D is neutralino and $D^\pm$ is chargino, here $s_D=1/2$. There exists a vast literature here, see for example, \cite{MSSMdark}.

\bu The second example  is given by inert doublet model (IDM)  \cite{inert}. In notations \cite{inert1}, this model contains one standard Higgs doublet $\phi_S$, responsible for electroweak symmetry breaking and generation of fermions and gauge bosons masses as in the Standard Model (SM), and another scalar doublet $\phi_D$, which doesn't receive  vacuum expectation value  and doesn't couple to fermions. Four degrees of freedom of  the Higgs doublet $\phi_S$ are as in the SM: three Goldstone modes  and one  mode becomes the Higgs boson  $h$.  All the components of the  second  scalar doublet $\phi_D$ are realized as massive scalar $D$-particles: two charged $D^\pm$ and two neutral ones $D$ and $D^A$. By construction, they possess a conserved multiplicative quantum number -- D-parity, the lightest particle among them is considered as a candidate for DM particle.  In our notations  the lightest D-particle  is  $D$. Here $s_D=0$.

Detailed limitations for masses of DM-particles from cosmology and collider physics are discussed in many papers (see e.g. \cite{MSSMdark}, \cite{limpap}). Below we have in mind   $M_D\le  70$~GeV   and $M_{D^\pm}>80$~GeV.

We consider production of such $D$-particles at $e^+e^-$ linear collider (e.g. at ILC) with electron beam energy $E_e\equiv \sqrt{s}/2$, provided that the 
family of $D$-particles with smaller weight than $E_e$ is settled by $D$, $D^\pm$ and no more than one neutral (for example $D^A$).
All  processes below we treat as basic reactions, $e^+e^-\to D^+D^-$ or\linebreak[4] $e^+e^-\to DD^A$ with subsequent decay of $D^\pm$ or $D^A$.

The problems which should be solved in the discussed experiments are following.\\
(1) To observe process with production of $D$-particles.\\
(2) To evaluate mass of DMP $M_D$ and masses of other $D$-particles.\\
(3) To evaluate spin of $D$-particles $s_D$.\\
(4) To obtain some additional information about  interactions of $D$-particles.

\section{Two type of $\pmb D$-particles, $\pmb D$ and $\pmb{D^\pm}$.}\label{sectmain}

In this section we consider the case when the set of $D$-particles in the energy region of ILC is settled by  $D^\pm$  and  $D$. In the considered models the only interaction that couples $D$ and $D^\pm$ with SM matter is  gauge interaction
 \be
 D^\pm D W^\mp. \label{DD+int}
 \ee
Therefore, $D^\pm\to DW^\pm$ with either on shell or off shell $W$'s is the only decay channel for $D^\pm$.

\subsection{Production, decay, signature}

We suggest to use  $D^+D^-$ production at ILC:
\be
e^+e^-\to D^+D^-\to DDW^+W^-\,.\label{mainpr}
\ee
The lab system here id c.m.s. for $e^+e^-$. In this system energies, $\gamma$-factors and velocities of $D^\pm$ are
 \be
 E_\pm=E_e,\;\; \gamma_\pm=E_e/M_{D^\pm}, \;\; \beta_\pm= \sqrt{1-M_{D^\pm}^2/E_e^2}.\label{cmkin}
 \ee

Neglecting quantity $1/4-\sin^2\theta_W^2$ (describing $\gamma-Z$ interference) the cross section of this process reads
\bea
&\sigma(e^+e^-\to D^+D^-)= &\nonumber\\
&=\fr{2\pi\alpha^2}{3s}\beta(3\!-\!\beta^2)\!\!
\left(1\!+\!\fr{R^{(1/2)}_Zs^2}{(s\!-\!M_Z^2)^2}\right)\;at\; \; s_D\!=\!\fr{1}{2};&\label{crsecferm}\\
&=\fr{\pi\alpha^2}{3s}\beta^3\left(1\!+\!\fr{R^{(0)}_Zs^2}{(s\!-\!M_Z^2)^2}\right)
\;\; at\;\; s_D=0;&\label{crsecscal}\\
&R_Z^{(1/2)}=\fr{1}{16\sin^4 (2\theta_W)}, \quad R_Z^{(0)}=\fr{\cot^2(2\theta_W)}{4\sin^2 (2\theta_W)}.&\nonumber
\eea
(In different models for DMP relative value of $Z$  contributions $R_Z^{(s)}$ can differs from these values by simple numerical factor). These cross sections are of the same order as $\sigma(e^+e^-\to\mu\mu)$ (which is given by  \eqref{crsecferm} at $\beta=1$). That are huge cross sections for ILC.

\bu {\bf If $\pmb{M_{D^\pm}-M_D>M_W}$}, the produced $D^\pm$ decays to $DW^\pm$ with mass shell $W$ only. We suggest to observe  the following final state systems:\\
\bu Two dijets from $q\bar{q}$ decay of $W^+$ and $W^-$, with effective mass $M_W$. For this channel the cross section is  $\sim \left[0.676^2\approx 0.45\right]\cdot\sigma(e^+e^-\to D^+D^-)$.\\
\bu One dijet from $q\bar{q}$ decay of $W^+$ or $W^-$ plus $\mu$ or $e$ from $\mu\nu$ or $e\nu$ or $\tau\nu\to \mu\nu\nu\nu$ or $\tau\nu\to e\nu\nu\nu$ decay of $W^-$ or $W^+$. For this channel the cross section is $\left[2\cdot 0.676\cdot2\cdot (1+0.17)0.108\approx 0.33\right]\! \cdot\sigma(e^+e^-\!\to\! D^+D^-)$
(here 0.17 is a fraction of $\mu$ or $e$ from the decay of $\tau$).

Typical event will have large missing energy carried out by neutral and stable $D$-particles, the observed momenta are strongly a-collinear, the missing  momentum is large and its direction is far from $e^+e^-$ collision axis.

The background  final state with the same kinematics can be produced in SM only if missed energy and $p_\bot$ are carried off by  neutrino(s).  The value of corresponding cross section is at least by one electroweak  coupling squared $g^2/4\pi$ or $g^{\prime 2}/4\pi$ less than $\sigma(e^+e^-\to\mu\mu)$ with $g^2/4\pi\sim g^{\prime 2}/4\pi\sim \alpha$. Therefore, cross sections for such background processes are about  two orders less than the cross section of the process under discussion. The same estimate is valid for all reactions, considered below.

\bu {\bf If $\pmb{M_{D^\pm}-M_D<M_W}$}, the only mode for charged $D^\pm$ decays is\fn{In the description of reaction we denote by $W$ and $Z$ both $W$ and $W^*$, $Z$ and $Z^*$. Here $W^*$ and $Z^*$ mean $q\bar{q}$ dijet or dilepton system with  quantum numbers of $W$ or $Z$ but with  lower effective mass.}
$DW^{*+}$. The structure of the final state and the value of cross sections are the same as in previous case with a single difference -- dijet effective mass $M^*$ is now not peaked around resonance value $M_W$ but distributed in some interval below $M_{D^\pm}-M_D$. The form of this distribution depends on the spin of $D$-particles $s_D$.

\bu The signature for the process in both cases is:
\bear{c}\boxed{\mbox{\begin{minipage}{0.4\textwidth}
Two dijets or one dijet plus $e$ or $\mu$ with large missing
 energy and large a-collinearity, with cross section $\sim\sigma(e^+e^-\to\mu^+\mu^-)$. Typically these dijets (or dijet and lepton) move in the opposite hemispheres.\end{minipage}
}}
\eear{sign1}

\subsection{Parameters of $\pmb{D^\pm}$ and $\pmb D$}

We denote
\be
\Delta(s;s_1,s_2)\!=\!\sqrt{s^2\!+\!s_1^2\!+\!s_2^2\!-\!2ss_1\!-\!2ss_2\!-\!2s_1s_2}.\label{tria}
\ee

{\bf The case $\pmb{M_{D^\pm}>M_{D^0}+M_W}$}.
In the rest frame of $D^\pm$ we deal with 2-particle decay $D^\pm\to DW^\pm$. The energy and momentum of $W^\pm$ in this frame are
 \be
\!E_W^r\!=\!\fr{M_{D^\pm}^2\!+\!M_{W}^2\!-\!M_{D}^2}{2M_{D^\pm}},\;\; p^r\!=\!\fr{\Delta(M_{D^\pm}^2,M_W^2,M_{D}^2)}{2M_{D^\pm}}.\label{rkin}
\ee
The effective mass of dijet is $M_W$ with accuracy to $W$ width.

Denoting  $W$ escape angle in $D^+$ rest frame relative to the direction of $D^+$ motion in the lab system by $\theta$ and $c=\cos\theta$ we have energy of $W^+$ in the lab system
 $$
 E_W^L=\gamma_\pm(E_W^r+c\beta_\pm p^r).
 $$
Therefore, energies of dijets W's are distributed within the interval
\be
\left(E_{(-)}\!=\!\gamma_\pm(E_W^r\!-\!\beta_\pm p^r)\,,\;E_{(+)}\!=\!\gamma_\pm(E_W^r\!+\!\beta_\pm p^r)\right).\label{EP}
 \ee

{\bf Masses}. {\it The end point values  $E_{(\pm)}$ give two equations for evaluation of masses $D^\pm$ and $D^0$.}

In particular, it is useful to  note that
 \be
 E_{(-)}E_{(+)}\!=\!\gamma_\pm^2 M_W^2+(p^r)^2,\;\;  E_{(-)}\!+\!E_{(+)}\!=\!2\gamma_\pm E^r_W.
 \label{Epprod}
\ee
Therefore at large enough electron energy (at $\gamma\gg 1$),
 \bear{c}
M_{D^\pm}^2\approx E_e^2M_W^2/[ E_{(-)}E_{(+)}], \\[2mm]
M_D^2=M_W^2+ \fr{E_e-E_{(-)}-E_{(+)}}{E_e}\,M_{D^\pm}^2\,.
\eear{heM}
At finite $\gamma$ the exact equations are more complex.

The accuracy of this procedure  is determined by the accuracy of measurement of dijet energy together with its effective mass and by a width of $D^\pm$ (if the latter is large). In particular, at $s_D=0$ the  decay $D^\pm\to DW^\pm$ width is
\be
\Gamma =\fr{\alpha}{2\sin^2\theta_W}\cdot \fr{(p^r)^3}{M_W^2}\,.\label{opendec+}
 \ee
The $\Gamma/M_{D^\pm}$ ratio is below 0.1 at $M_{D^\pm}\le 500$~GeV.

The distribution of these dijets in energy is uniform, $dN(E)
\propto dE$ since there is no correlation between escape angle of $W$ in the rest frame of $D^\pm$ and production angle of $D^\pm$. When the width of $D^\pm$ is not small, this distribution become non-uniform near the end points. The measuring of fine structure of this distribution  near the end point will give the total $D^\pm$ width at least roughly.

\bu {\bf If $\pmb{M_{D^\pm}>M_{D^0}+M_W}$}, the only decay channel is  $D^+\to DW^{*+}$.
All discussed above results are valid for each separate value $M^*$, with evident change in all equations $M_W\to M^*$.

The energy and $M^*$ distributions for each pair of dijets are independent from each other.

The masses of $M_{D^\pm}$ and $M_D$ are evaluated in this case for each measured value of $M^*$ even with the best accuracy than in the previous case since in this case proper width of $D^\pm$ is low enough.

{\bf Spin}. After evaluation of $M_{D^\pm}$, the cross section of $e^+e^-\to D^+D^-$ process is calculated precisely in QED for each $s_D$.
The cross section for $s_D=0$ \eqref{crsecscal} is more than four times less than that for $s_D=1/2$ \eqref{crsecferm}. This big difference allows conclude definitely about spin of $D$-particles.

{\bf Other properties for $\pmb{s_D=1/2}$}.  If $D^\pm$ are fermions, their spins are correlated to each other and to longitudinal polarization of collided electrons. Observation  of process $e^+e^-\to D^+D^-\to D^0D^0jj\ell+\nu$'s allows to find the sign of charge of dijet $W=q\bar{q}$ in each separate case. It allows to study the charge and polarization asymmetries for accessing  of more detail properties of $D$-particles (e.g. ratio of $D^+D^-\gamma$ to $D^+D^-Z$ couplings).

\section{Two  neutral $\pmb D$-particles  } \label{sectDD1}

Here we consider the case when the set of $D$-particles in the energy region of ILC consists of single charged $D$-particle $D^\pm$ and two   neutral $D$-particles with identical CP, $D$ and $D_1$ with $M_{D^\pm}>M_{D_1}>M_D$, where $D$ can be  DMP while $D_1$ can disappear during lifetime of the Universe.  In the discussion below we neglect an exotic case of approximate degeneracy between $D$ and $D_1$.

\bu Let the only  interaction that couples $D$ or $D_1$ and $D^\pm$ with SM matter is the standard gauge interaction \eqref{DD+int}.
In this case process \eqref{mainpr} is supplemented by similar processes with one or two $D_1$ instead of $D$ in the final state:
 \be
e^+e^-\to D^+D^-\to \left\{ \begin{array}{cc} DDW^+W^-&(a)\,,\\
DD_1W^+W^-\,,&(b)\\
D_1D_1W^+W^-\,.&(c)
  \end{array}\right.\label{cascDD1}
 \ee
with the same signature \eqref{sign1}. The probability of the decay $D^\pm\to D_1W^\pm$ is lower than the probability of the  decay $D^\pm\to DW^\pm$ due to smaller final phase space volume in the first case. Therefore
 $\sigma(a)>\sigma(b)>\sigma(c)$.

The description of basic process $e^+e^-\to D^+D^-$ coincides with that presented above. Each decay\linebreak[4]  $D^\pm\to DW^\pm$ and  $D^\pm\to D_1W^\pm$ is described as above, the only new point is that the end points $E_{(-)}$  and $E_{(+)}$ for energy distribution of $W$'s from decay\linebreak[4] $D^\pm\to DW^\pm$ are given by equations \eqref{EP} while  the end points $E_{(-)}^1$  and $E_{(+)}^1$ for energy distribution of $W$'s from decay $D^\pm\to D_1W^\pm$ are given by the same equations but with the change $M_D$ to $M_{D_1}$. It is evident that $E_{(-)}<E_{(-)}^1<E_{(+)}^1<E_{(+)}$. Therefore in this case the same procedure as above allows to obtain masses $M_{D^\pm}$ and $M_D$.

To evaluate $M_{D_1}$ let us remind that for each type of decay the energy distribution of $W$ in the lab. system is uniform. The energy distribution of $W$ in the lab. system is the sum of two uniform distributions with the described above end points. The end points  $E_{(-)}^1,\,E_{(+)}^1$ are marked by the steps in the density of event energy distribution. They can be used for evaluation of $M_{D_1}$.

The distribution of dijets in the effective mass can be different. If all masses are peaked around $M_W$, it means that $M_{D^\pm}-M_{D}>M_{D^\pm}-M_{D_1}>M_W$. If there are dijets with effective mass $M_W$ and those with lower effective mass, the former appear from $D^\pm \to DW^\pm$ decay and the latter -- from $D^\pm \to D_1W^\pm$ decay. They can be used for evaluation of $M_D$ and $M_{D_1}$ separately. If the effective masses of all dijets are below $M_W$, we have\linebreak[4] $M_W>M_{D^\pm}-M_{D}>M_{D^\pm}-M_{D_1}$. In this case mentioned steps in the dijet energy distribution at each $M^*$ will be added by steps in the distribution in $M^*$ for evaluation of $M_D$ and $M_{D_1}$.

Both channels of $D^\pm$ decay must be taken into account at  comparison  of the measured  cross sections with \eqref{crsecferm}, \eqref{crsecscal} for evaluation of spin $s_D$.

\bu In some models {\bf interaction $\pmb{D_1Dh}$ is allowed} additionally. We discuss this opportunity assuming that $h$ is anticipated Higgs boson of SM with mass about 120 GeV, very small width and dominant $b\bar{b}$ decay channel. Modifications of description for another properties of $h$, observed at LHC before operations at LC, is simple.
In this case reactions (\ref{cascDD1}(b)) and (\ref{cascDD1}(c)) are modified as
 \be
e^+e^-\!\to\! D^+D^-\!\to\! \left\{ \begin{array}{cc}
DD_1W^+W^-\!\to\! DDW^+W^-h\,,&(b)\\
\!D_1D_1W^+W^-\!\to\! DDW^+W^-hh.&(c)
  \end{array}\right.\label{cascDD1h}
 \ee
The signature \eqref{sign1} is naturally added by one or two $b\bar{b}$ dijets from the Higgs decay. Using of $b$ tagging allows to distinguish these dijets from $W$ dijets which is necessary for the evaluation of masses. If the ratio $(M_{D_1}-M_D)/M_h$ is small enough, the $D_1$ time of life can be so large that one can observe it in the detector.

\section{Two neutral  $\pmb D$-particles with opposite CP-parity }

In the IDM  together with $D$ and $D^\pm$ the one more neutral scalar particle $D^A$ exists (with mass $M_{D^A}>M_D$)  \cite{inert1}. Here CP parities of $D$ or $D^A$ cannot be defined separately since they do not interact to fermions, but their relative parity is fixed, they have opposite CP-parities. It results in  the gauge interaction $D^ADZ$ and instability $D^A$ via the decay $D^A\to DZ$ (in addition to gauge interaction $D^AD^+W^-$ similar to that for $D$, $DD^+W^-$). Similar particles, interaction and production channel are possible also in some models for fermionic Dark Matter.

So, we discuss now the case when additional neutral $D$-particle $D^A$. In this discussion we neglect case of approximate degeneracy between $D$ and $D^A$.\\

\bu{\bf If $\pmb {M_{D^A}+M_D<2E_e,\; M_{D^A}<M_{D^\pm}}$},
The most important  new process is
\be
e^+e^-\to DD^A\to DDZ.\label{DAprod}
\ee
Instead of \eqref{cmkin}, $D^A$ energy, $\gamma$-factor and velocity are
 \bear{c}
E_A=\fr{4E_e^2+M_{D^A}^2-M_D^2}{4E_e},\\[3mm] \gamma_A=\fr{E_{D^A}}{M_{D^A}},\quad \beta_A=\fr{\Delta(4E_e^2, M_{D^A}^2, M_D^2)}{4E_e E_{D^A}}\,.
\eear{Aprodcm}

In the IDM (at $s_D=0$) the cross section
\be
\sigma(e^+e^-\to DD^A)=\fr{\pi\alpha^2sR^{(0)}_Z}{3(s-M_Z^2)^2}\beta_A^3\fr{E_A}{2E_e-E_A}\,.\label{Acrsec}
\ee
This cross section is of the same order of value as $\sigma(e^+e^-\to\mu^+\mu^-)$.

The signature of this process is similar to that given by eq.~\eqref{sign1}:
\be\boxed{\mbox{\begin{minipage}{0.4\textwidth}
One $q\bar{q}$ dijet or $e^+e^-$ or $\mu^+\mu^-$ pair with identical effective mass and energy distributions and with large missing $E_\bot$.\end{minipage}
}}
\label{sign2}
\ee

If $M_{D^A}-M_D>M_Z$, observable final state  is $Z$, which is seen as hadronic dijet or $e^+e^-$ or $\mu^+\mu^-$ with effective mass equal to $M_Z$. Energy distribution of this $Z$ is given by equations similar to \eqref{rkin},\eqref{EP}. End points of this distribution allow to evaluate masses $M_{D}$ and $M_{D^A}$.

If  $M_{D^A}-M_D<M_Z$,  observable final state is  $Z^*$,  which is seen as hadronic dijet or $e^+e^-$ or $\mu^+\mu^-$ with identical spectra of effective mass. For each value of this effective mass energy distribution of this $Z^*$ is given by equations similar to \eqref{rkin},\eqref{EP}. End points of this distribution allow to evaluate masses $M_{D}$ and $M_{D^A}$.

After the study of process \eqref{DAprod} one must to study process \eqref{mainpr} and  cascade reactions
 \be
e^+e^-\!\!\to\! D^+D^-\!\!\to\! \left\{\! \begin{array}{ll} DDW^+W^-\,,&(a)\\
  D^AW^\pm DW^\mp\!\to\! DD W^+W^-Z,&(b)\\ D^AW^+D^AW^-\!\to\! DD W^+W^-ZZ.&(c)
  \end{array}\right.\label{cascD}
 \ee
The decay $D^\pm \to D^AW^\pm$ is described by the same equation as the decay $D^\pm \to DW^\pm$. Its probability is lower than that for decay $D^\pm \to DW^\pm$ due to smaller final phase space volume. Therefore,  $\sigma(a)>\sigma(b)>\sigma(c)$.

The signature of the process (\ref{cascD}(a)), just as the process (\ref{cascD}(b), (c)) for  invisible decays of $Z$, is given by \eqref{sign1}. Just as in the analysis of sect.~\ref{sectDD1}, the study of end points in the energy distribution of $W$ allows to obtain $M_D$ and $M_{D^\pm}$ and steps in this distribution can be used to verify value of $M_{D^A}$, obtained via a study of process with the signature \eqref{sign2}.

The signature of processes (\ref{cascD}(b)) and (\ref{cascD}(c)) for visible decays of $Z$ is similar to  \eqref{sign1} with adding of dijet or $\ell^+\ell^-$ pairs which represent $Z$ or $Z^*$. If one can distinguish dijets from $Z$ and those from $W$, the energy distribution of $W$ in these processes can be used to enhance data massive for evaluation of masses, discussed above.

Summation of these 3 cross sections gives total cross section of $e^+e^-\to D^+D^-$ process, which value gives us the value of spin $s_D$.\\

\bu {\bf If $\pmb {M_{D^A}+M_D<2E_e,\; M_{D^A}>M_{D^\pm}}$},
the analysis of sect.~\ref{sectmain} is valid completely for the final states with signature \eqref{sign1} -- reaction $e^+e^-\to D^+D^-$.

The second series of processes is $e^+e^-\to DD^A$ with two channels of $D^A$ decay and different signatures
 \be
 e^+e^-\!\!\to\! DD^A\!\to\!\left\{\!\begin{array}{cl}
  DDZ\,,&(a) \\
   DD^\pm W^\mp\to DDW^+W^-\,.&(b)
  \end{array}\right.\label{badpr}
  \ee

The process (\ref{badpr}(a)) is the process \eqref{DAprod}. It can be analysed just as it was discussed earlier. The process (\ref{badpr}(b)) is a cascade process. It can be eliminated from mass analysis of process \eqref{mainpr} by using the fact that in difference with the process \eqref{mainpr} the observable decay products of this process move typically in one hemisphere.\\

\section{Summary}

We present simple and robust method for discovery candidates for DM particles and evaluation of its mass and spin (with evaluation of mass and spin of some other particles with the same $D$-parity)  at ILC/CLIC. 

These particles will be discovered via observation of processes with signature \eqref{sign1}, \eqref{sign2} and with cross section of the order of $\sigma(e^+e^-\to\mu^+\mu^-$, which is huge for LC. The masses of these particles will be obtained via measuring the details of energy distribution of dijets (representing $W^\pm$ or $Z$) near the end points. The cross section  measurement  of process with signature \eqref{sign1}, \eqref{sign2}  and similar signature for the derivative processes with cascade decay allows to determine the spin $s_D$ of considered candidate for DM particle by comparison with simple SM calculation (the cross sections for $s_D=1/2$ is approximately 4 times larger than that for $s_D=0$).

One of proposed processes, $e^+e^-\to D^+D^-$, was considered earlier in respect of discovery of  neutralino as DMP, etc. (see e.g.  \cite{ILCuse}). However, I never saw such  approach for simultaneous evaluation of masses and spins of $D$-particles irrespective to details of model.
The advantages of presented approach are following:\\
{\it 1. The cross section of each suggested process is substantial part of the total cross section of $e^+e^-$ annihilation at considered energy (typically up to tens percents).\\
2. The signature is clear.\\
3. The background is very small (typically,  $\sim 1$~\% from the observable effect).\\
4. The kinematics is very simple, allowing reliable extraction of quantities under interest from the data.}\\
We have no idea how one use similar approach at hadronic colliders.

I am thankful D.Yu. Ivanov  and K.A. Kanishev for discussions.
The work  was  supported by grants RFBR 08-02-00334-a, NSh-3810.2010.2 and Program of Dept. of Phys. Sc. RAS "Experimental and theoretical studies of fundamental interactions related to LHC."

\end{document}